\documentclass{vldb}

\usepackage{epsfig,endnotes}
\usepackage{array}
\usepackage[utf8]{inputenc}
\usepackage{balance}
\usepackage{totcount}
\usepackage{etoolbox}
\usepackage{xcolor}
\usepackage{pagecolor}
\usepackage{enumitem}
\usepackage{url}
\usepackage{amsmath}
\usepackage{algorithm}
\usepackage[noend]{algpseudocode}
\algrenewcommand\algorithmicrequire{\textbf{Input:}}
\algrenewcommand\algorithmicensure{\textbf{Output:}}
\newtoggle{shownotes}
\toggletrue{shownotes}
\newtotcounter{numnotes}

\newtoggle{showtodos}
\toggletrue{showtodos}
\newtotcounter{numtodos}

\newcommand{\checkforcrud}{
  \ifboolexpr{test{\ifnumcomp{\totvalue{numnotes}}{>}{0}} or
              test{\ifnumcomp{\totvalue{numtodos}}{>}{0}}}{
    \pagecolor{red!20}
  }{}
}

\newcommand{\smallitem}[1]{\vspace{0.3em}\noindent\textbf{#1}}
\newcommand{\smallitembot}{\vspace{0.5em}\noindent}

\newcommand{\system}{Anna}
\newcommand{\indy}{Anna v0}

\newcolumntype{P}[1]{>{\centering\arraybackslash}m{#1}}

\vldbTitle{}
\vldbAuthors{}
\vldbDOI{}
\vldbVolume{}
\vldbNumber{}
\vldbYear{}

\begin{document}

\title{Eliminating Boundaries in Cloud Storage with \system{}}

%for single author (just remove % characters)
\author{
\alignauthor
    Chenggang Wu, Vikram Sreekanti, Joseph M. Hellerstein\\
    \affaddr{UC Berkeley}\\
    \email{\{cgwu, vikrams, hellerstein\}@berkeley.edu}
} % end author

\maketitle

% Use the following at camera-ready time to suppress page numbers.
% Comment it out when you first submit the paper for review.
% \thispagestyle{empty}

% Uncomment this line to check if any comments, notes, or TODOs are remaining in the text.
% The body of the paper will turn red if any are left. If it stays white, there are none.
\checkforcrud{}

\abstract{
In this paper, we describe how we extended a distributed key-value store called \system{} into an elastic, multi-tier service for the cloud. 
In its extended form, \system{} is designed to overcome the narrow cost-performance limitations typical of current cloud storage systems.
We describe three key aspects of \system{}'s new design: multi-master selective replication of hot keys, a vertical tiering of storage layers with different cost-performance tradeoffs, and horizontal elasticity of each tier to add and remove nodes in response to load dynamics. 
\system{}'s policy engine uses these mechanisms to balance service-level  objectives around cost, latency and fault tolerance. 
Experimental results explore the behavior of \system{}'s mechanisms and policy, exhibiting orders of magnitude efficiency improvements over both commodity cloud KVS services and research systems.
}

\section{Introduction}\label{sec:intro}

As public infrastructure cloud providers have matured in the last decade, the number of storage services they offer has soared.
Popular cloud providers like Amazon Web Services (AWS)~\cite{aws}, Microsoft Azure~\cite{azure}, and Google Cloud Platform (GCP)~\cite{gcp} each have at least seven storage options.
These services span the spectrum of cost-performance tradeoffs: AWS ElastiCache, for example, is an expensive, memory-speed service, while AWS Glacier is extremely high-latency and low-cost.
In between, there are a variety of services such as the Elastic Block Store (EBS), the Elastic File System (EFS), and the Simple Storage Service (S3). 
Azure and GCP both offer a similar range of storage solutions.

Each one of these services is tuned to a unique point in that design space, making it well-suited to certain performance goals. 
Application developers, however, typically deal with a non-uniform \emph{distribution} of performance requirements. 
For example, many applications generate a skewed access distribution, in which some data is ``hot'' while other data is ``cold''. 
This is why traditional storage is assembled hierarchically: hot data is kept in fast, expensive cache while cold data is kept in slow, cheap storage.
These access distributions have become more complex in modern settings, because they can change dramatically over time. 
Realistic workloads spike by orders of magnitude, and hot sets shift and resize. 
These large-scale variations in workload motivate an ``elastic'' service design, but most cloud storage services today are inelastic and unable to respond to these dynamics.

The narrow performance goals of cloud storage services result in poor cost-performance tradeoffs for applications.
To improve performance, developers often take matters into their own hands by addressing storage limitations in custom application logic.
This introduces significant complexity and increases the likelihood of application-level errors.
Developers are inhibited by two key types of boundaries when building applications with non-uniform workload distributions:

\smallitem{Cost-Performance Boundaries}.
Each of the systems discussed above---ElastiCache, EBS, S3, etc.---offers a different, \emph{fixed} tradeoff of cost, capacity, latency, and bandwidth. 
These tradeoffs echo traditional memory hierarchies built from RAM, flash, and magnetic disk arrays. 
To balance performance and cost, data should ideally move adaptively across storage tiers, matching workload skew and shifting hotspots.
However, current cloud services are largely unaware of each other, so software developers and DevOps engineers must cobble together ad hoc memory hierarchies.
Applications must explicitly move and track data and requests across storage system boundaries in their business logic.
This task is further complicated by the heterogeneity of storage services in terms of deployment, APIs, and consistency guarantees. 
For example, single-replica systems like ElastiCache are linearizable, while replicated systems like DynamoDB are eventually consistent.

\smallitem{Static Deployment Boundaries}.
Cloud providers offer very few truly elastic storage services; most such systems have hard boundaries on the number and type of nodes deployed.
In AWS for example, high performance tiers like ElastiCache are surprisingly inelastic, requiring system administrators to allocate and deallocate instances manually. 
Two of the lower storage tiers---S3 and DynamoDB---are elastic, but are insufficient for many needs.
S3 autoscales to match data volume but ignores workload; it is designed for ``cold'' storage, offering good bandwidth but high latency. 
DynamoDB offers workload-based autoscaling but is prohibitively expensive to scale to a memory-speed service.
This motivates the use of ElastiCache over DynamoDB, which again requires an administrator to monitor load and usage statistics, and manually adjust resource allocation.

\smallitembot

In an earlier paper, we presented the initial architecture of a key-value storage system called \system{}~\cite{anna}. 
The focus of the initial paper was on a design to provide excellent performance across orders of magnitude in scale. 
In this work, we extend \system{} to remove its cost-performance and static deployment boundaries, enabling \system{} to dynamically adjust configuration and match resources to workloads.

While our previous work's evaluation focused on raw performance, here we are also interested in \emph{efficiency}: the ratio of performance to cost.
For various cost points, we show that \system{} outperforms in-memory systems like AWS ElastiCache and Masstree~\cite{mao2012cache} by up to an order of magnitude. 
\system{} also outperforms elastic databases like DynamoDB by more than two orders of magnitude in efficiency.

In Section~\ref{sec:background}, we briefly describe the contributions of our prior work on \system{}~\cite{anna} and preview our approach to making the system adapt across boundaries.
In Section~\ref{sec:design}, we describe the mechanisms that \system{} uses to respond to mixed and changing workloads. 
Section~\ref{sec:arch} introduces the architecture of \system{} including the implementation of these mechanisms, and Section~\ref{sec:policy} describes \system{}'s policy engine.
In Section~\ref{sec:eval}, we present an evaluation of \system{}'s mechanisms and policies, and we describe how they fare in comparison to the state of the art. 
Section~\ref{sec:related} discusses related work, and we conclude with future work in Section~\ref{sec:conclusion}.

In the remainder of this paper, we use AWS as the public cloud provider underlying \system{}. 
The design principles and lessons learned here are naturally transferable to other cloud providers with similar offerings.
\section{Background}\label{sec:background}

The first paper on \system{}~\cite{anna} presented a distributed key-value store based on a fully shared-nothing, thread-per-core architecture with background gossip across cores and nodes.  
\system{} threads have no shared data structures in memory beyond message queues, enabling each core to spend most of its time doing useful work.
Experiments on high-contention workloads showed Anna spending over 90\% of its compute cycles  serving  put  and  get  requests,  while  state-of-the-art, competing  systems  were  achieving  less  than  10\%.  
The vast majority of the other systems’ time was spent trying to execute atomic processor instructions on shared data structures.  
As a result, \system{} outperformed the competition by orders of magnitude at many scale points.
Relative to the state-of-the-art distributed KVSes, Anna’s initial design also enabled an unprecedented richness  of coordination-free  consistency  levels.   
The basic insight was that the full variety of coordination-free consistency  and  transactional  isolation  levels  taxonomized  by Bailis et al.~\cite{Bailis:2013:HAT:2732232.2732237} can be achieved by the monotone composition of simple lattice structures, as suggested by Conway, et al.~\cite{Conway:2012:LLD:2391229.2391230}. 
The original paper maps out a wide range of key-value and NoSQL systems against Bailis’ taxonomy of consistency levels.

The first version of \system{} focused on performing well at both single-node and distributed scales.
This showed that eventual consistency combined with a coordination-free shared nothing architecture makes data management easy in the face of deployment changes and also hinted at the potential to remove deployment boundaries.
However, the initial architecture lacked the mechanisms to monitor and respond to usage and workloads.
Another notable weakness of the initial work was its need to aggressively replicate the entire database across the main memory of many machines to achieve high performance. 
This gave the system an unattractive cost-performance tradeoff and made its resource allocation very rigid.
As a result, although a benchmark-beater, \system{}’s first version suffered from the problems highlighted above: it was expensive and inflexible for large datasets with non-uniform access distributions.

\pagebreak
\subsection{Anna Without Boundaries}\label{sec:background-adpativity}

In this paper, we extend the initial version of \system{} to  span  the  cost-performance design space more flexibly, enabling it to adapt dynamically to workload variation in a cloud-native setting.
The architecture presented here removes the cost-performance and static deployment boundaries discussed in Section~\ref{sec:intro}.
To that end, we add three key mechanisms: (1) horizontal elasticity to adaptively scale deployments; 
(2) vertical data movement in a storage hierarchy to reduce cost by demoting cold data to cheap storage; 
and (3) multi-master  selective  replication of hot keys across nodes and cores to efficiently scale request handling for non-uniform access patterns.
The resulting architecture is simplified by deploying a single storage kernel across many tiers, by entirely avoiding coordination, and by reusing the storage engine to store and manipulate system metadata.
The additions to Anna described in this work enable system operators to specify high-level goals such as fault tolerance and cost-performance objectives, without needing to manually configure the number of nodes and the replication factors of keys.  
A new policy engine automatically responds to workload shifts using the mechanisms mentioned above to meet these service-level objectives (SLOs).
\section{Distributions and Mechanisms}\label{sec:design}

In this section, we first classify and describe common workload distributions across data and time.
We then discuss the mechanisms that \system{} uses to respond to the workload properties and changes.

We believe that an ideal cloud storage service should gracefully adapt to three aspects of workload distributions and their dynamics in time:

\smallitem{A. Volume}. As overall workload grows, the aggregate throughput of the system must grow.
During growth periods, the system should automatically increase resource allocation and thereby cost. 
When workload decreases, resource usage and cost should decrease correspondingly as well.

\smallitem{B. Skewness}. Even at a fixed volume, skewness of access distributions can affect performance dramatically.
A highly skewed workload will make many requests to a small subset of keys.
A uniform workload of similar volume will make a few requests to each key.
Different skews warrant different responses, to ensure that the resources devoted to serving each key are proportional to its popularity.

\smallitem{C. Shifting Hotspots}. Workloads that are static in both skew and volume can still exhibit changes in distribution over time: hot data may become cold and vice versa.
The system must be able to detect changes in workload hotspots and respond accordingly by prioritizing data in the new hot set and demoting data in the old one.
\smallitembot

We address these three workload variations with three mechanisms in \system{}, which we describe next.

\smallitem{1. Horizontal Elasticity}.
In order to adapt to variation in workload volume, each storage tier in \system{} must scale elastically and independently, both in terms of storage and request handling.
\system{} needs the storage capacity of many nodes to store large amounts of data, and it needs the compute and networking capacity of many nodes to serve large numbers of requests.
This is accomplished by partitioning (sharding) data across all the nodes in a given tier.
When workload volume increases, \system{} can respond by automatically adding nodes and repartitioning a subset of data. 
When the volume decreases, \system{} can remove nodes and repartition data among the remainders.

\smallitem{2. Multi-Master Selective Replication}.
When workloads are highly skewed, simply adding shards to the system will not alleviate pressure.
The small hot set will be concentrated on a few nodes that will be receiving a large majority of the requests, while the remaining nodes lie idle. 
The only solution is to replicate the hot set onto many machines. 
However, we do not want to repeat the mistakes of our first iteration of~\system{}'s design, replicating cold keys as well---this simply wastes space and increases overhead. Instead, replication must be selective, with hot keys replicated more than cold keys.
Thus, \system{} must accurately track which data is hot and which is cold, and the replication factors and current replica locations for each key.

\smallitem{3. Vertical Tiering}.
As in a traditional memory hierarchy, hot data should reside in a fast, memory-speed tier for efficient access; significant cost savings are available by demoting data that is \emph{not} frequently accessed to cold storage.
Again, \system{} must correctly classify hot and cold data in order to promote or demote appropriately.
While the previous two mechanisms are aimed at improving performance, this one primarily attempts to minimize cost without compromising performance.

\subsection{Summary}

\begin{table}
\centering
{\small
\begin{tabular} { |c|c| }
    \hline
    \textbf{Workload Dynamics} & \textbf{Relevant Mechanisms} \\
    \hline
    Volume & Elasticity \\
    \hline
    Skew & Replication, Tiering \\
    \hline
    Hotspot & Replication, Tiering \\
    \hline
\end{tabular}
\caption{The mechanisms used by \system{} to deal with various aspects of workload distributions.}
\label{table:mechanism}
}
\end{table}

Table~\ref{table:mechanism} shows which mechanisms respond to which properties of workload distributions.
There is a direct mapping between an increase (or decrease) in volume---with other factors held constant---and a requirement to elastically add (or remove) nodes.
Changes in workload skew require a response to the new hot set size via promotion or demotion, as well as appropriate selective replication.
Similarly, a change in hotspot location requires correct promotion and demotion across tiers, in addition to shifts in per-key replication factors.
We describe how \system{} implements each one of these mechanisms in Sections~\ref{sec:arch} and~\ref{sec:policy}.
In Section~\ref{sec:eval}, we evaluate how well \system{} responds to these dynamics.
\section{Anna Architecture}\label{sec:arch}

In this section, we introduce \system{}'s architecture and illustrate how the mechanisms discussed in Section~\ref{sec:design} are implemented.
We present an overview of the core subsystems and then discuss each component in turn.
As mentioned in Section~\ref{sec:intro}, \system{} is built on AWS components.
In our initial implementation and evaluation, we validate this architecture over two storage tiers: one providing RAM cost-performance and another providing flash disk cost-performance.
\system{}'s memory tier stores data in RAM attached to AWS EC2 nodes.
The flash tier leverages the Elastic Block Store (EBS), a fault-tolerant block storage service that masquerades as a mounted disk volume on an EC2 node. 
There is nothing intrinsic in our choice of layers. 
We could easily add a third layer (e.g., S3) and a fourth (e.g., Glacier), but demoting data to cold storage in these tiers operates on much longer timescales that are beyond the scope of this work.

\subsection{Overview}\label{sec:arch-overview}

\begin{figure}[t]
  \centering
    \includegraphics[width=0.48\textwidth]{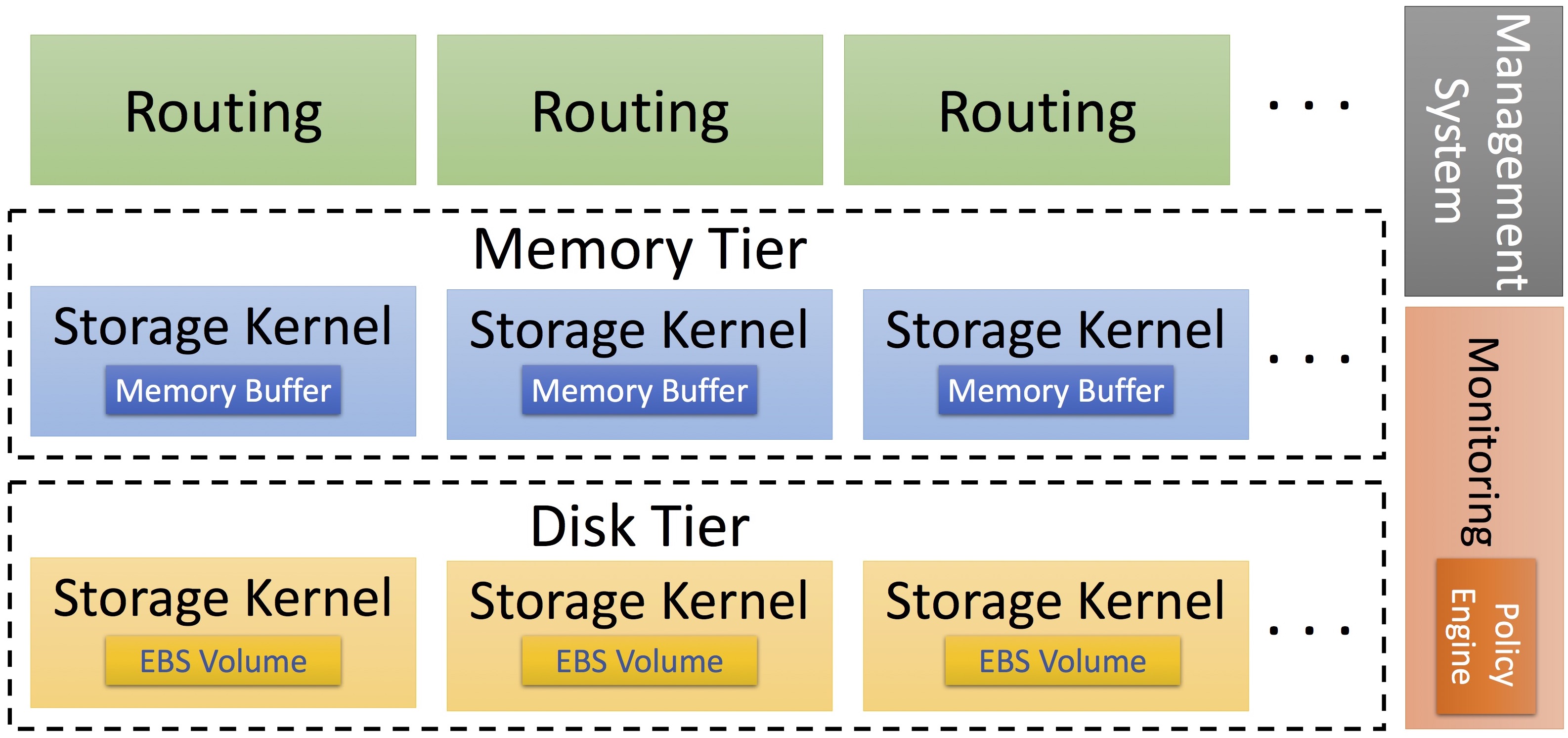}
  \caption{The \system{} architecture.}
  \label{fig:arch}
\end{figure}

Figure~\ref{fig:arch} presents an overview of \system{}, with each rectangle representing a node.
In the original \system{} paper~\cite{anna}, we described an extremely performant, coordination-free key-value store that provided a rich variety of consistency levels.
In that work, we demonstrated how a KVS could scale from multicore to distributed settings while gracefully tolerating the natural messaging delays that arise in distributed systems.
To enable the mechanisms described in Section~\ref{sec:design}, we first extended the storage kernel to support multiple storage media and then designed three new subsystems: a monitoring system/policy engine, a routing service, and a cluster management system.
Each subsystem is bootstrapped on top of the key-value storage component in \system{}, storing and modifying their metadata as keys and values in the system.

The monitoring system and policy engine are the internal services responsible for responding to workload dynamics and meeting SLOs.
Importantly, these services are stateless and thus are not concerned with fault tolerance and scaling; they rely on the storage service for these features.

The routing service is a stateless client-facing API that provides a stable abstraction above the internal dynamics of the system.
The resource allocation of each tier may be in flux---and whole tiers may be added or removed from the system at workload extremes---but clients are isolated from these changes.
The routing service consistently returns a correct endpoint that will answer client requests.
Finally, the cluster management system is another stateless service that executes resource allocation changes based on decisions reached by the policy engine.

\subsection{Storage System}\label{sec:arch-storage-system}

\begin{figure}[t]
  \centering
    \includegraphics[width=0.5\textwidth]{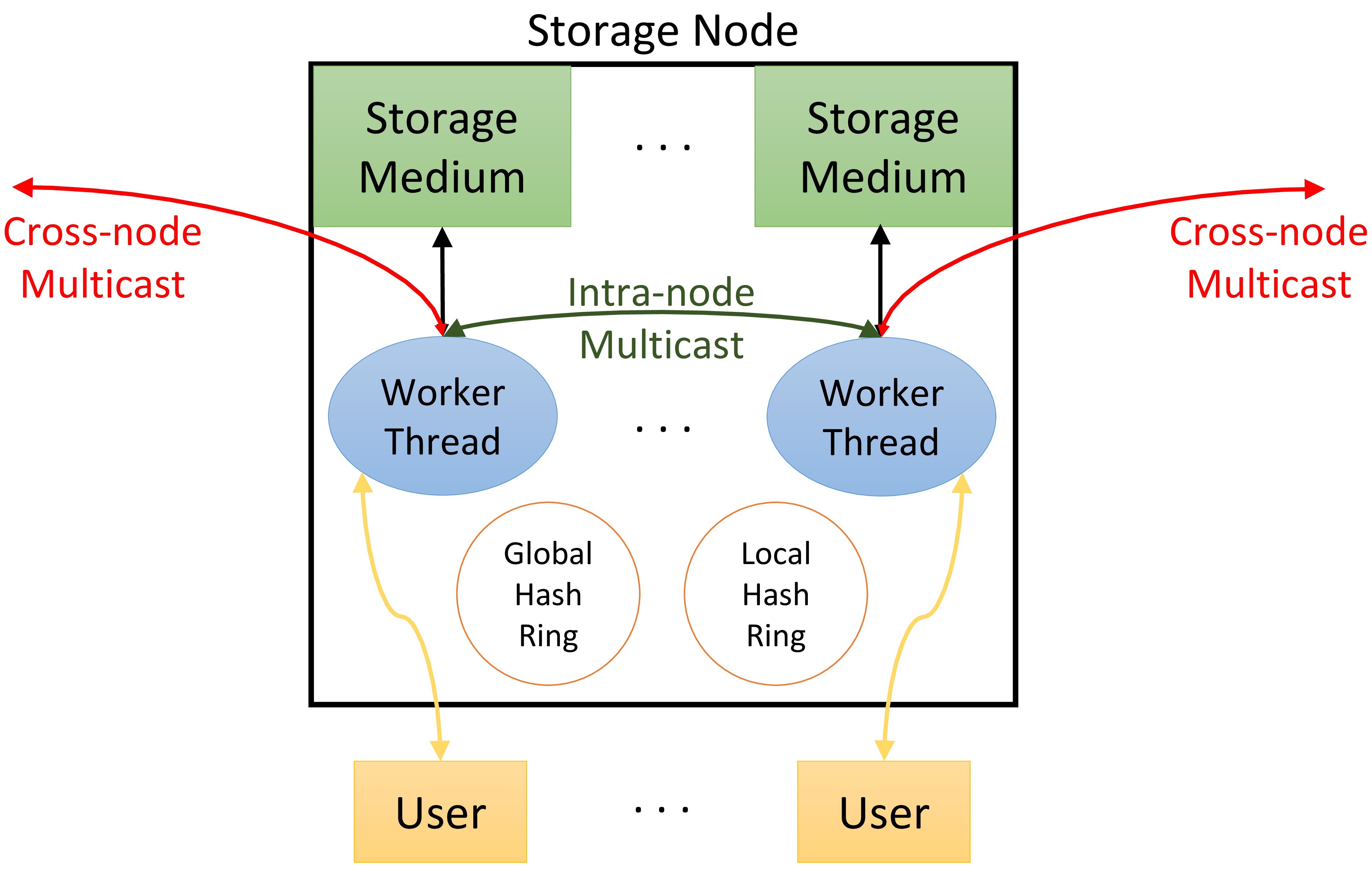}
  \caption{The architecture of a storage node.}
  \label{fig:storage}
\end{figure}

Figure~\ref{fig:storage} shows the architecture of an \system{} storage node.
Each node has many worker threads, and each thread interacts with a thread-local storage medium (a memory buffer or disk volume), processes client requests, and sends/receives multicasts to/from other \system{} workers.

The \system{} storage kernel is deployed across many storage tiers.
The only difference between tiers is the procedure for translating data for persistence (serialization/deserialization, a.k.a.\ ``serde'').
Memory-tier nodes read from and write to local memory buffers, while disk-tier nodes serialize data into files that are stored on EBS volumes.
\system{}'s uniformity across storage tiers makes adding additional tiers very simple: we set the serde mode and adjust the number of worker threads based on the underlying hardware.
For instance, the total number of threads for memory nodes matches the number of CPU cores to fully utilize computing resources and to avoid costly preemption of threads. 
However, in other storage tiers where the performance bottleneck lies in serializing the key-value pairs to and from persistent storage, the optimal strategy for resource allocation is different.
Our EBS tier allocates 4$\times$ as many threads per node (4) as we have physical CPUs (1).

\system{} uses consistent hashing~\cite{Karger:1997:CHR:258533.258660} to partition and replicate keys.
For performance and fault tolerance (discussed further in Sections~\ref{sec:policy} and~\ref{sec:eval}), each key may be replicated onto many nodes in each tier and multiple threads in each node.
Following the model of early distributed hash tables, we use virtual nodes~\cite{rao2003load} in our consistent hashing algorithm. 
Each physical node (or thread) handles traffic for many virtual nodes (or threads) on the hash ring to ensure an even  distribution.
Virtual nodes also enable us to add heterogeneous nodes in the future by allocating more virtual nodes to more powerful physical machines.

In the following section, we present a brief overview of the storage kernel's design points that enable it to achieve high-performance coordination-free execution and replica consistency.
This overview is a brief summary of the initial design of \system{} presented in \cite{anna}.

\subsubsection{Storage Kernel} \label{sec:arch-anna overview}

Recent work has demonstrated that shared-memory coordination mechanisms like locking and atomic ``lock-free'' instructions slow down low-level memory access performance on a single node by orders of magnitude~\cite{latchfreecidr17, Boyd-wickizer_non-scalablelocks}.
Across nodes, consensus algorithms are well-known to cause dramatic latency and availability problems~\cite{Brewer:2010:CFT:1835698.1835701, pacelc, birman2009toward}.
\system{}'s coordination-free execution model avoids these issues entirely in pursuit of excellent performance and scalability. 
It gives each worker thread on every node a private memory buffer to store the data it manages. 
Data is multi-mastered: each thread or node processes both reads and writes locally regardless of replication.
Each thread periodically runs a background task to multicast (``gossip'') recent updates to other workers that maintain replicas of these keys.
This shared-nothing, asynchronous messaging scheme eliminates thread synchronization and asynchronously resolves conflicting updates to replicas.
The resulting code exploits multi-core parallelism within a single machine and smoothly scales out across distributed nodes. Our earlier work shows dramatic benefits from this design, including record performance based on extremely high (90\%) CPU utilization in useful work with low processor cache miss rates.

While \system{} eliminates contention, consistency becomes tricky: the same set of updates may arrive at different replicas in different orders. 
Naïvely applying these updates can cause replicas to diverge and lead to inconsistent state.
Another contribution of~\cite{anna} is achieving a wide range of consistency models by encapsulating state into monotone compositions of simple CRDT-style~\cite{crtpshapiro} lattices, inspired by the Bloom language~\cite{Conway:2012:LLD:2391229.2391230}.
Lattices tolerate message reordering and duplication while guaranteeing eventual convergence of replicas. 
By default, \system{} stores data in last-writer-wins lattices, which resolve divergent updates by picking the update with the most recent timestamp.
However, \system{}'s lattices can be composed to offer the full range of coordination-free consistency guarantees including causal consistency, item cut isolation, and read-committed transactions~\cite{Bailis:2013:HAT:2732232.2732237}. 

\subsection{Metadata Management} \label{sec:arch-metadata}

\system{} requires maintaining certain metadata to efficiently support mechanisms discussed in Section~\ref{sec:design} and help the policy engine adapt to changing workloads.
In this section, we introduce the types of metadata managed by \system{} and how they are stored and used by various system components.

\subsubsection{Types of Metadata}

\system{} manages three distinct kinds of metadata.
First, every storage tier has two \emph{hash rings}. 
A global hash ring, $G$, determines which nodes in a tier are responsible for storing each key.
A local hash ring, $L$, determines the set of worker threads \emph{within} a single node that are responsible for a key.
Second, each individual key $K$ has a \emph{replication vector} of the form $[< R_1, ... R_n>, <T_1, ... T_n>]$.
$R_i$ represents the number of nodes in tier $i$ storing key $K$, and $T_i$ represents the number of threads per node in tier $i$ storing key $K$.
During client-request handling and multicast, both hash rings and key $K$'s replication vector are required to determine the set of threads responsible for the key.
For every tier, $i$, that maintains a replica of $K$, we first hash $K$ against $G_i$ to determine which nodes are responsible for $K$.
We then look at $L_i$, tier $i$'s local hash ring to determine which worker threads are responsible for the key.

Lastly, \system{} also tracks monitoring statistics, such as the access frequency of each key and the storage consumption of each node.
This information is analyzed by the policy engine to trigger actions in response to variations in workload.
Currently, we store 16 bytes of metadata per key and about 10 KB of metadata per worker thread.

\subsubsection{Metadata Storage}

Clearly, the availability and consistency of metadata is as important as that of regular data---otherwise, \system{} would be unable to determine a key's location or get an accurate estimate of workload characteristics and the system's resource usage.
In many systems~\cite{hadoop, storm, heron, greenplum}, metadata is enmeshed in the implementation of ``master nodes'' or stateful services like ZooKeeper~\cite{hunt2010zookeeper}.
\system{} simply stores metadata in the storage system.
Our metadata automatically derives all the benefits of our storage system, including performance guarantees, fault tolerance, and consistency. 
\system{} employs last-writer-wins consistency to resolve conflicts among metadata replicas.
Due to the eventual consistency model, worker threads may have stale views of hash rings and replication vectors.
This can cause threads to disagree on the location of a key and can potentially cause multiple rounds of request redirection.
However, since the metadata will eventually converge, threads will agree on the key's location, and requests will reach the correct destination.
Note that multicast is performed every few seconds, while cluster state changes on the order of minutes, so cluster state metadata is guaranteed to converge before it undergoes further changes.
In summary, storing metadata in \system{} both simplifies system design by reducing external software dependencies and improves performance by relaxing the required consistency model.
\subsubsection{Enabling Mechanisms}
\label{sec:arch-metadata-mechanisms}

Interestingly, manipulating two of these types of metadata (hash rings and replication vectors) is the key to enabling the mechanisms described earlier in Section~\ref{sec:design}.
In this section, we discuss \emph{only} the implementation of each mechanism.
When and why each action is executed is a matter of policy and will differ based on system configuration and workload characteristics---we save this discussion for Section~\ref{sec:policy}.

\smallitem{Elasticity}.
When a new node joins a storage tier, it queries the storage system to retrieve the hash ring, updates the ring to include itself, and broadcasts its presence to all nodes in the system---storage, monitoring, and routing.
Each existing node updates its copy of the hash ring, determines if it stores any keys that the new node is now responsible for, and gossips those keys to the new node.
Similarly, when a node is removed, it removes itself from the hash ring and broadcasts its departure to all nodes.
It then determines which nodes are now responsible for its data and gossips its keys to those nodes.
Once all data has been broadcast, the node goes offline and its resources are deallocated.

Key migration overheads can be significant (see Section~\ref{sec:eval-dynamic-workload}).
To address this challenge, \system{} interleaves key migration with client request handling to prevent system downtime. 
This is possible due to \system{}'s support for coordination-free consistency: The client may retrieve stale data during the key migration phase, but it can maintain a client-side cache and merge future retrieved results with the cached value.
\system{}'s lattice-based conflict resolution guarantees that the state of the cached data is monotonically growing.

\smallitem{Selective Replication \& Cross-Tier Data Movement}.
Both these mechanisms are implemented via updates to replication vectors.
Each key in our two-tier implementation has a default replication vector of the form $[<1, k>, <1, 1>]$, meaning that it has one memory tier replica and $k$ EBS-tier replicas.
Here, $k$ is the number of replica faults per key the administrator is willing to tolerate (discussed further in Section~\ref{sec:arch-fault-tolerance} and~\ref{sec:policy}).
By default, keys are not replicated across threads within a single node.
\system{} induces cross-tier data movement by simply manipulating metadata. It increments the replication factor of one tier and decrements that of the other tier; as a result, gossip migrates data across tiers.
Similarly, selective replication is achieved by adjusting the replication factor in each tier, under the fault tolerance constraint.
After updating the replication vector, \system{} synchronizes the metadata across replicas via asynchronous multicast.

\subsection{Monitoring System \& Policy Engine} \label{sec:arch-monitoring-system}

\begin{figure}[t]
  \centering
    \includegraphics[width=0.40\textwidth]{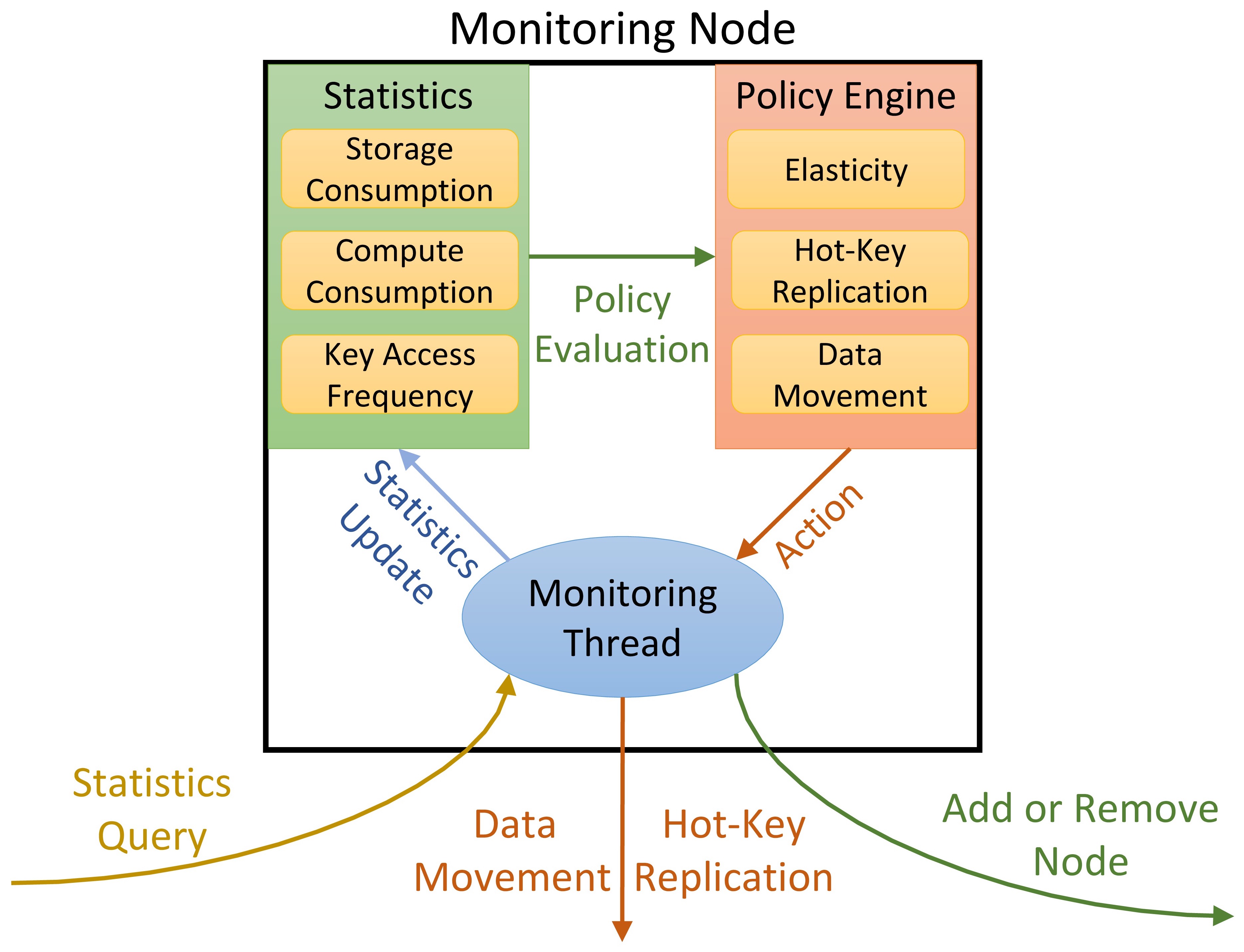}
  \caption{Monitoring node architecture.}
  \label{fig:monitoring}
\end{figure}

In this section, we discuss the design of the monitoring system and how it interacts with the policy engine. 
As shown in Figure~\ref{fig:monitoring}, each monitoring node has a monitoring thread, a statistics buffer, and a policy engine. 
The monitoring thread is stateless and periodically retrieves the stored statistics from the storage engine and triggers the policy engine.
The policy engine in turn analyzes these statistics and issues actions to meet its SLOs.
\system{} currently supports three types of actions: \emph{elasticity change}, \emph{hot-key replication}, and \emph{cross-tier data movement}.
The implementation of these actions is covered above, in Section~\ref{sec:arch-metadata-mechanisms}.
We discuss when each of these actions is triggered and describe the end-to-end policy algorithm in Section~\ref{sec:policy}.

\subsection{Routing Service}\label{sec:arch-routing-service}

The routing service isolates clients from the underlying storage system: A client asks where to find a key and is returned the set of all valid addresses for that key.
\system{}'s routing service only maintains soft state. 
Each routing node caches the storage tiers' hash rings and key replication vector metadata to respond to the clients' key address requests.
If a key has any memory-tier replicas, the routing service only returns memory-tier addresses to maximize performance.
The client caches these addresses locally to reduce request latency and load on the routing service.

When a client's cached address set becomes invalid because of a change in cluster configuration, a storage server will tell the client to invalidate cache entries for keys that have moved.
The routing service will refresh its cached cluster state and give the client a new set of addresses, which will again be cached until they are invalidated.

\subsection{Cluster Management}\label{sec:arch-implementation-elasticity}

\system{} uses Kubernetes~\cite{kubernetes} as a cluster management tool.
Kubernetes is responsible for allocating and deallocating nodes, ensuring that nodes are alive, and rebooting failed nodes.
An \system{} deployment has four kinds of nodes: storage nodes, routing nodes, monitoring nodes, and a single, stateless ``cluster management'' node described below.

A ``pod'' is the atomic unit of a Kubernetes application and is a collection of one or more Docker~\cite{docker} containers.
All containers within a pod have access to the same resources but are isolated from each other.
Each node in \system{} is instantiated in a separate Kubernetes pod, and each pod contains only one instance of a \system{} node.
Storage system and routing service pods are pinned on separate EC2 instances for resource isolation purposes.
The monitoring system is less resource intensive and can tolerate preemption, so it is not isolated.
Finally, \system{} maintains a singleton cluster management pod, whose role is to issue requests to add or remove nodes to the Kubernetes cluster.
A simple, stateless Python server in this pod receives REST requests from the policy engine and uses bash scripts to add or remove nodes.

\subsection{Fault Tolerance}\label{sec:arch-fault-tolerance}

\system{} guarantes $k$-fault tolerance by ensuring $k + 1$ replicas are live at all times.
The choice of $k$ determines a trade-off between resilience and cost.
The $k+1$ replicas of each key can be spread across tiers arbitrarily, according to hotness.

When a storage node fails, other nodes detect the failure via a timeout and remove the node from the hash ring.
When such a timeout happens, \system{} automatically repartitions data using the updated hash ring.
The cluster management pod then issues a request to spawn a new node, which enters the join protocol discussed in Section~\ref{sec:arch-metadata-mechanisms}.

\system{} does not rely on the persistence of EBS volumes for fault tolerance in the disk tier.
Similar to nodes in the memory tier, these nodes lose their state when they crash---this is desirable because it allows all tiers to be symmetric, regardless of the durability of the underlying storage medium.

Both routing nodes and monitoring nodes only store soft state and do not require any recovery mechanisms.
If a routing node fails, it queries other routing nodes for up-to-date cluster information, and if a monitoring node fails, it retrieves system statistics from the storage service.

When the cluster management pod fails, Kubernetes automatically revives it. 
No recovery is necessary as it does not manage any state.
The state of the cluster will not change while the pod is down since it is the actor responsible for modifying resource allocation.
As a result, the policy engine will re-detect any issue requiring an elasticity change before the crash and re-issue the request upon revival.

\smallitembot

In summary, \system{} consists of a stateful storage kernel that is partitioned and selectively replicated for performance and fault tolerance with multi-master updates.
Every other component is either stateless and optionally caches soft state that is easily recreated.
As a result, the only single point of failure in \system{} is the Kubernetes master. 
Kubernetes offers high-availability features to mitigate this problem~\cite{kubernetes-ha}.
We also note that Kubernetes is not integral to the design of \system{}; we rely on it primarily to reduce the engineering burden of mundane tasks such as receiving heartbeats, allocating VMs, and deploying containers.
\section{Policy Engine} \label{sec:policy}

\system{} supports three kinds of SLOs: an average request latency ($L_{obj}$) in milliseconds, a cost budget ($B$) in dollars/hour, and a fault tolerance ($k$) in number of replicas.
The fault tolerance indicates the allowed number of replica failures, $k$.
The latency objective, $L_{obj}$, is the average expected request latency.
The budget, $B$, is the maximum cost per hour that will be spent on \system{}. 

As discussed in Section~\ref{sec:arch-fault-tolerance}, \system{} ensures there will never be fewer than $k + 1$ replicas of each key to achieve the fault tolerance goal.
The latency objective and cost budget goals, however, are conflicting.
The cheapest configuration of \system{} is to have $k + 1$ EBS nodes and 1 memory node (for metadata).
Clearly, this configuration will not be very performant.
If we increase performance by adding memory nodes to the system, we might exceed our budget. 
Conversely, if we strictly enforce the budget, we might not be able to achieve the latency objective.

The \system{} administrator specifies only \emph{one} of the two goals.
If a latency SLO is specified, \system{} minimizes the cost while meeting the latency.
If the budget is specified instead, \system{} uses no more than $\$B$ per hour while maximizing performance.

In Sections~\ref{sec:policy-data-movement},~\ref{sec:policy-hot-key-replication}, and~\ref{sec:policy-elasticity}, we describe heuristics to trigger each policy action---data movement, hot key replication, and elasticity.
In Section~\ref{sec:policy-algorithm}, we present \system{}'s complete policy algorithm, which combines these heuristics to achieve the SLO. 
Throughout this section, we represent each key's replication vector as $[<R_M, R_E>, <T_M, T_E>]$ since our initial prototype only uses two tiers---$M$ for memory and $E$ for EBS.
We have included pseudocode for the algorithms in this section in the Appendix.

\subsection{Cross-Tier Data Movement} \label{sec:policy-data-movement}

\system{}'s policy engine uses its monitoring statistics to calculate how frequently each key was accessed in the past $T$ seconds, where $T$ is an internal parameter.
If a key's access frequency exceeds a configurable threshold, $P$, and  all replicas currently reside in the EBS tier, \system{} promotes a single replica to the memory tier.
If the key's access frequency falls below a separate internal threshold, $D$, and the key has one or more memory replicas, all replicas are demoted to the EBS tier. 
The EBS replication factor is set to $k+1$, and the local replication factors are restored to 1.
Note that in \system{}, all metadata is stored in the memory tier, is never demoted, and has a constant replication factor.
If the aggregate storage capacity of a tier is full, \system{} adds nodes (Section~\ref{sec:policy-elasticity}) to increase capacity before performing data movement.
If the budget does not allow for more nodes, \system{} employs a least-recently used caching policy to demote keys.

\subsection{Hot-Key Replication} \label{sec:policy-hot-key-replication}

When the access frequency of a key stored in the memory tier increases, hot-key replication increases the number of memory-tier replicas of that key.
In our initial implementation, we configure only the memory tier to replicate hot keys. 
Because the EBS tier is not intended to be as performant, a hot key in that tier will first be promoted to the memory tier before being replicated.
This policy will likely vary for a different storage hierarchy.

The policy engine classifies a key as ``hot'' if its access frequency exceeds an internal threshold, $H$, which is $s$ standard deviations above the mean access frequency. 
Because Anna is a shared-nothing system, we can replicate hot keys both across cores in a single node and across nodes.
Replicating across nodes seems preferable, because network ports are a typical bottleneck in distributed system, so replicating across nodes multiplies the aggregate network bandwidth to the key. 
However, replicating across cores within a node can also be beneficial, as we will see in Section~\ref{sec:eval-replica-placement}. 
Therefore, hot keys are \emph{first} replicated across more nodes before being replicated across threads within a node.

The policy engine computes the target replication factor, $R_{M\_ideal}$, using the ratio between the observed latency for the key and the latency objective. 
Cross-node replication is only possible if the current number of memory replicas, $R_M$,  is less than the number of memory-tier nodes in the cluster, $N_M$.
If so, we increment the key's memory replication factor to $min(R_{M\_ideal}, N_M)$.
Otherwise, we increment the key's \emph{local} replication factor on memory-tier machines up to the maximum number of worker threads ($N_{T\_memory})$ using the same ratio.
Finally, if the access frequency of a previously-hot key drops below a threshold, $L$, its replication vector is restored to the default: $R_M$, $T_M$, and $T_E$ are all set to 1 and $R_E$ is set to $k$.

\subsection{Elasticity}\label{sec:policy-elasticity}

\smallitem{Node Addition}.
\system{} adds nodes when there is insufficient storage or compute capacity.
When a tier has insufficient storage capacity, the policy engine computes the number of nodes required based on data size, subject to cost constraints, and instructs the cluster management service to allocate new nodes to that tier.

Node addition due to insufficient compute capacity only happens in the memory tier because the EBS tier is not designed for performance.
Compute pressure on the EBS tier is alleviated by promoting data to the memory tier since a memory node can support 15$\times$ the requests at 4$\times$ the cost.
The policy engine uses the ratio between the observed latency and the latency objective to compute the number of memory nodes to add.
This ratio is bounded by a system parameter, $c$, to avoid overly aggressive allocation.

\smallitem{Node Removal}.
\system{} requires a minimum of one memory node (for system metadata) and $k+1$ EBS nodes (to meet the $k$-fault SLO when all data is demoted).
The policy engine respects these lower bounds.
We first check if any key's replication factor will exceed the total number of storage nodes in any tier after node removal. 
Those keys' replication factors are decremented to match the number of nodes at each tier before the nodes are removed.
\system{} currently only scales down the memory tier based on compute consumption and not based on storage consumption.
This is because selective replication can significantly increase compute consumption without increasing storage consumption.
Nonetheless, this may lead to wasteful spending under adversarial workloads; we elaborate in the next section.

\smallitem{Grace Periods}.
When resource allocation is modified, data is redistributed across each tier, briefly increasing request latency (see Section~\ref{sec:eval-dynamic-workload}). 
Due to this increase, as well as data location changes, key access frequency decreases.
To prevent over-correction during key redistribution, we apply a grace period to allow the system to stabilize.
Key demotion, hot-key replication, and elasticity actions are all delayed till after the grace period.

\subsection{End-to-End Policy} \label{sec:policy-algorithm}

In this section, we discuss how \system{}'s policy engine combines the above heuristics to meet its SLOs.
If the average storage consumption of \emph{all} nodes in a particular tier has violated configurable upper or lower thresholds ($S_{upper}$ and $S_{lower}$), nodes are added or removed respectively.
We then invoke the data movement heuristic from Section~\ref{sec:policy-data-movement} to promote and demote data across tiers. 
Next, the policy engine checks the average latency reported by clients.
If the latency exceeds a fraction, $f_{upper}$ (defaulting to 0.75), of the latency SLO and the memory tier's compute consumption exceeds a threshold, $C_{upper}$, nodes are added to the memory tier.
However, if not all nodes are occupied, hot keys are replicated in the memory tier, as per Section~\ref{sec:policy-hot-key-replication}.
Finally, if the observed latency is a fraction, $f_{lower}$ (defaulting to 0.5), below the objective and the compute occupancy is below $C_{lower}$, we invoke the node removal heuristic to check if nodes can be removed to save cost. 

The compute threshold, $C_{upper}$, is set to $0.20$.
Consistent with our previous work~\cite{anna}, each storage node saturates its network bandwidth well before its compute capacity.
We use the compute occupancy as a proxy metric for the saturation of the underlying network connection.
This threshold varies significantly based on the hardware configuration; we found that 20\% was optimal for our experimental setup (see Section~\ref{sec:eval}).

\subsubsection{Discussion}

\smallitem{Storage Node Saturation}.
There are two possible causes for saturation. 
If all nodes are busy processing client requests, \system{} must add more nodes to alleviate the load. 
Performing hot-key replication is not productive: Since all nodes are busy, replicating hot keys to a busy node will, in fact, decrease performance due to additional gossip overhead. 
The other cause is a skewed access distribution in which most client requests are sent to a small set of nodes serving the hot keys while most nodes are free.
The optimal solution is to replicate the hot keys onto unsaturated nodes.
If we add nodes to the cluster, the hot keys' replication factors will not change, and clients will continue to query the few nodes storing those keys.
Meanwhile, the newly added nodes will idle.
As discussed in Section~\ref{sec:policy-algorithm} (Line 8 and 10 of Algorithm 5 in the Appendix), \system{}'s policy engine is able to differentiate the two causes for node saturation and take the appropriate action.

\smallitem{Policy Limitations}.
There are cases in which our policy engine fails to meet the latency objective and/or wastes money.
Due to current cloud infrastructure limitations, for example, it takes about five minutes to allocate a new node.
An adversary could easily abuse this limitation.
A short workload spike to trigger elasticity, followed by an immediate decrease would lead \system{} to allocate unnecessary nodes.
These nodes will be under-utilized, but will only be removed if the observed latency drops below $f_{lower} * L_{obj}$.
Unfortunately, removing this constraint would make \system{} susceptible to reducing resource allocation during network outages, which is also undesirable.
We discuss potential solutions to these issues in future work.

\smallitem{Knobs}.
There are a small number of configuration variables mentioned in this section, which are summarized in Table~\ref{table:knobs}.
We distinguish variables that are part of the external SLO Spec from the internal parameters of our current policy.
In our evaluation, our parameters were tuned by hand to match the characteristics of the AWS services we use. 
There has been interesting work recently on autotuning database system configuration knobs~\cite{VanAken:2017:ADM:3035918.3064029}; our setting has many fewer knobs than those systems.
As an alternative to auto-tuning our current knobs, we are exploring the idea of replacing the current threshold-based policy entirely with a dynamic Reinforcement Learning policy that maps directly and dynamically from performance metrics to decisions about system configuration changes. 
These changes to the policy engine are easy to implement, but tuning the policy is beyond the scope of this paper: It involves extensive empirical work on multiple deployment configurations.

\begin{table}
\centering
{\small
\begin{tabular}{ |P{1.1cm}|P{1.75cm}|P{2.5cm}|P{1.5cm}| }
     \hline
     \textbf{Variable Name} & \textbf{Meaning} & \textbf{Default Value} & \textbf{Type} \\
     \hline
     $L_{obj}$ & Latency Objective & 2.5ms & SLO Spec \\\hline
     $B$ & Cost Budget & N/A (user-specified) & SLO Spec \\\hline
     $k$ & Fault Tolerance & 2 & SLO Spec \\\hline
     $T$ & Monitoring report period & 30 seconds & Policy Knob \\\hline
     $H$ & Key hotness threshold & 3 standard deviations above the mean key access frequency & Policy Knob \\\hline
     $L$ & Key coldness threshold & The mean key access frequency & Policy Knob \\\hline     
     $P$ & Key promotion threshold & 2 accesses in $T$ seconds & Policy Knob \\\hline
     [$S_{lower}$, $S_{upper}$] & Storage consumption thresholds & Memory: [0.3, 0.6] EBS: [0.5, 0.75] & Policy Knob \\\hline
     [$f_{lower}$, $f_{upper}$] & Latency thresholds & [0.5, 0.75] & Policy Knob \\\hline
     [$C_{lower}$, $C_{upper}$] & Compute occupancy thresholds & [0.05, 0.20] & Policy Knob\\\hline
     $c$ & Upper bound for latency ratio & 1.5 & Policy Knob \\\hline
\end{tabular}
\caption{A summary of all variables mentioned in Section~\ref{sec:policy}.}
\label{table:knobs} 
}
\end{table}
\section{Evaluation}\label{sec:eval}

In this section, we present an evaluation of \system{}.
We first explore the advantage of different replica placement strategies in Section~\ref{sec:eval-replica-placement}.
We then show the benefit of selective replication in Section~\ref{sec:eval-selective}.
We demonstrate \system{}'s ability to detect and adapt to variation in workload volume, skew, and hotspots in Section~\ref{sec:eval-dynamic-workload} and~\ref{sec:eval-hotspot}.
Finally, Section~\ref{sec:eval-trade-off} evaluates \system{}'s ability to trade off performance and cost according to its SLO.

When selecting the appropriate instance type, we measured the best combination of memory, CPU, and network bandwidth for an average workload; due to space constraints, we do not include an evaluation here.
\system{} uses \texttt{r4.2xlarge} instances for memory-tier nodes and \texttt{r4.large} instances for EBS-tier nodes.
Each node has 4 worker threads; at peak capacity they can handle a workload that saturates the network link of the node.
\texttt{r4.2xlarge} memory nodes have 61GB of memory, which is equally divided among all worker threads. 
Each thread in a EBS node has access to its own 64GB EBS volume.
In our experiments, \system{} uses two \texttt{m4.large} instances for the routing nodes and one \texttt{m4.large} instance for the monitoring node.
We include these nodes in all cost calculation below.
Unless otherwise specified, all experiments are run on a database with 1 million key-value pairs.
Keys and values are 8 bytes and 256KB long, respectively.
We set the $k$-fault tolerance goal to $k=2$; there are 3 total replicas of each key.
This leads to a total dataset size of about 750GB: $1M\ keys \times 3\ replicas \times 256KB\ values$.

Our workload is a YCSB-style read-modify-write of a single key chosen from a Zipfian distribution.
We adjust the Zipfian coefficient to create different contention levels---a higher coefficient means a more skewed workload.
The clients were run on \texttt{r4.16xlarge} machines, with 8 threads each.
Unless stated otherwise, experiments used 40 client machines for a total of 320 concurrent, single-threaded clients. 

\subsection{Replica Placement}\label{sec:eval-replica-placement}

We first compare the benefits of intra-node vs.\ cross-node replication; for brevity, no charts are shown for this topic.
On 12 memory-tier nodes, we run a highly skewed workload with the Zipfian coefficient set to 2.
With a \emph{single} replica per key, we observe a maximum throughput of just above 2,000 operations per second (ops).
In the case of cross-node replication, four \emph{nodes} each have one thread responsible for each replicated key; in the intra-node case, we have only one node with four \emph{threads} responsible for each key.
Cross-node replication improves performance by a factor of four to 8,000 ops, while intra-node replication only improves performance by a factor of two to 4,000 ops.
This is because the four threads on a single node all compete for the same network bandwidth, while the single threads on four separate nodes have access to four times the aggregate bandwidth.
Hence, as discussed in Section~\ref{sec:policy-hot-key-replication}, we prioritize cross-node replication over intra-node replication whenever possible but \emph{also} take advantage of intra-node replication.

\subsection{Selective Replication}\label{sec:eval-selective}

\begin{figure}[t]
  \centering
    \includegraphics[width=0.5\textwidth]{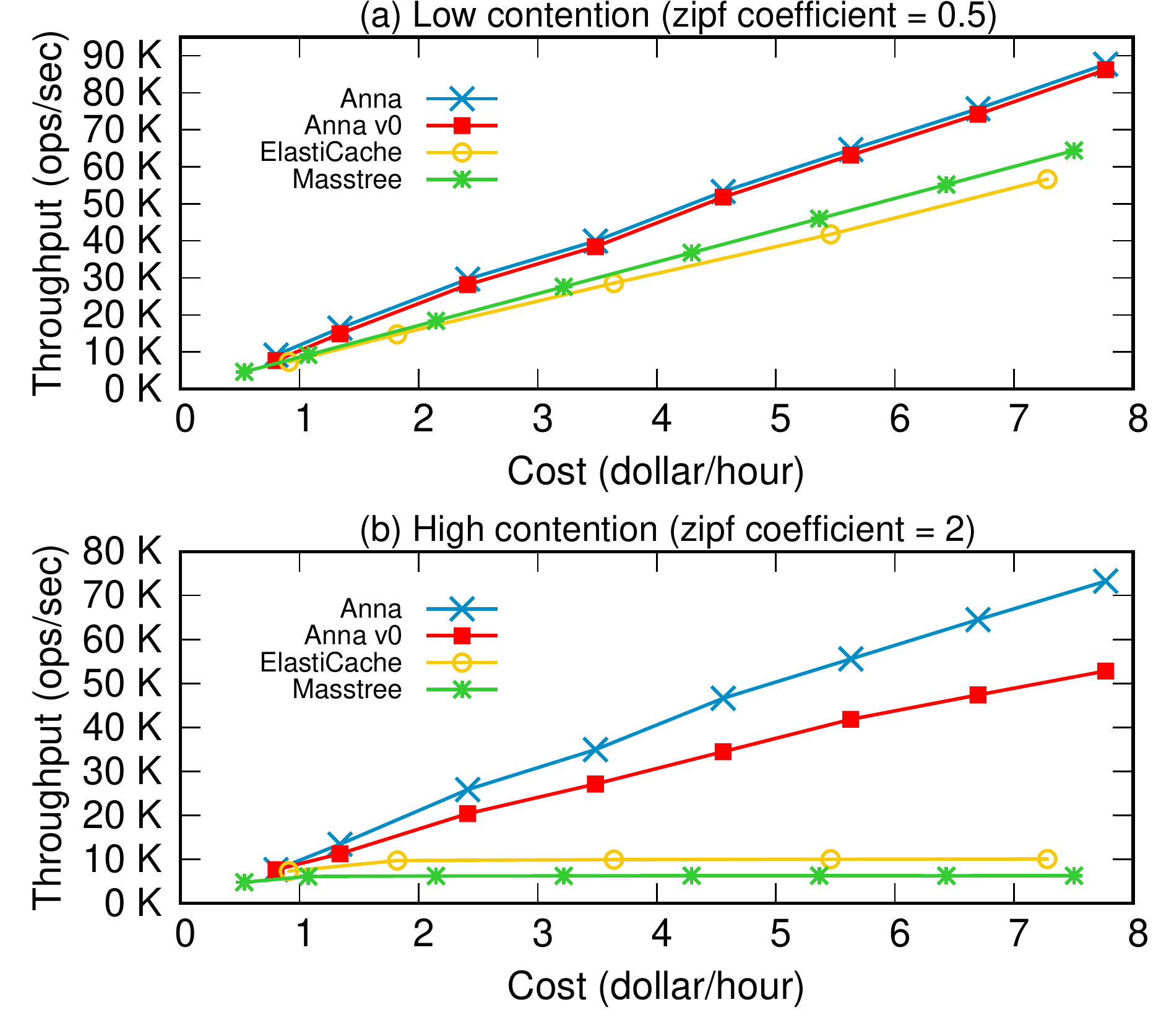}
  \caption{Cost-effectiveness comparison between \system{}, \indy{}, Elasticache, and Masstree.}
  \label{fig:comparison}
\end{figure}

A key weakness of our initial work~\cite{anna} (referred to as \indy{}) is that all keys are assigned a uniform replication factor. 
A poor choice of replication factor can lead to significant performance degradation.
Increasing the replication factor boosts performance for skewed workloads, as requests to hot keys can be processed in parallel on different replicas.
However, a uniform replication factor means that cold keys are \emph{also} replicated, which increases gossip overhead (slowing down the system) and storage utilization (making the system more expensive). By contrast, \system{} selectively replicates hot keys to achieve high performance, without paying a storage cost for replicating cold keys.

This experiment explores the benefits of selective replication by comparing \system{}'s memory-tier against \indy{}, AWS ElastiCache (using managed Memcached), and a leading research system, Masstree~\cite{mao2012cache}, at various cost points.
We hand-tune \indy{}'s single replication factor to the optimal value for each Zipfian setting and each cost point.
This experiment uses a database of 100,000 keys across all cost points; we use a smaller database since the data must fit on one node, corresponding to the minimum cost point.
We configure keys in \system{} to have a default replication factor of 1 since neither ElastiCache nor Masstree supports replication of any kind.
To measure the performance for a fixed price, we also disabled \system{}'s elasticity mechanism.

Figure~\ref{fig:comparison}(a) shows that \system{} consistently outperforms both Masstree and ElastiCache under low contention.
As discussed in our previous work, this is because \system{}'s thread-per-core coordination-free execution model efficiently exploits multi-core parallelism, while other systems suffer from thread synchronization overhead through the use of locks or atomic instructions.
Neither \system{} nor \indy{} replicates data in this experiment, so they deliver identical performance.

Under high contention (Figure~\ref{fig:comparison}(b)), \system{}'s throughput increases linearly with cost, while both ElastiCache and Masstree plateau.
\system{} selectively replicates hot keys across nodes and threads to spread the load, enabling this linear scaling; the other two systems do not have this capability.
\indy{} replicates the \emph{entire} database across all nodes.
While \indy{}'s performance scales, the absolute throughput is worse than \system{}'s because naively replicating the entire database increases multicast overhead for cold keys.
Furthermore, \indy{}'s storage consumption is significantly higher: At \$7.80/hour (14 memory nodes), \indy{}'s constant replication generates 13$\times$ the original data size, while \system{} incurs $<$1\% extra storage overhead.

\subsection{Dynamic Workload Skew \& Volume}\label{sec:eval-dynamic-workload}

\begin{figure}[t]
  \centering
    \includegraphics[width=0.5\textwidth]{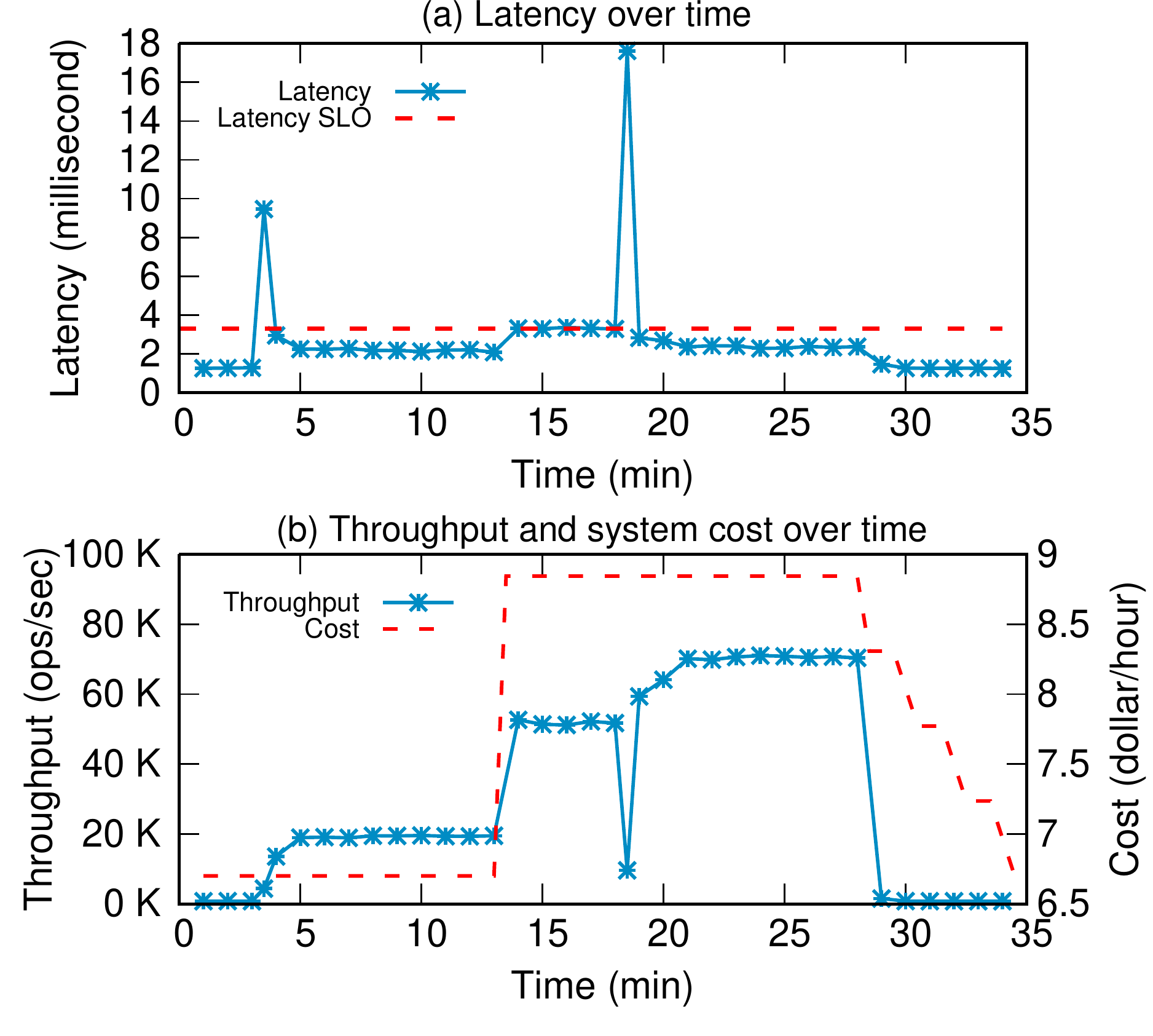}
  \caption{\system's response to changing workload.}
  \label{fig:dynamic-workload}
\end{figure}

We now combine selective replication and elasticity to react to changes in workload skew and volume.
In this experiment, we start with 12 memory-tier nodes and a latency objective of 3.3ms---about 33\% above our unsaturated latency.
All servers serve a light load at time 0.
At the 3-minute point, we start a high contention workload with a Zipfian coefficient of 2.
We see in Figure~\ref{fig:dynamic-workload}(a) that after a brief spike in latency, \system{} replicates the highly contended keys and meets the latency SLO (the dashed red line).
At minute 13, we reduce the Zipfian coefficient to 0.5 to switch to a low contention workload.
Simultaneously, we increase the load volume by a factor of 4.
Detecting these changes, the policy engine reduces the replication factors of the previously-hot keys.
It finds that all nodes are occupied with client requests and issues a request to add four more nodes to the cluster.
We see a corresponding increase in the system cost in Figure~\ref{fig:dynamic-workload}(b).

It takes 5 minutes for the new nodes to join the cluster.
Throughput increases to the saturation point of all nodes (the first plateau in Figure~\ref{fig:dynamic-workload}(b)), and the latency spikes to the SLO maximum from minutes 13 to 18.
At minute 18, the new nodes come online and trigger a round of data repartitioning, seen by the brief latency spike and throughput dip. 
\system{} then further increases throughput and meets the latency SLO.
At the 28-minute point, we reduce the load, and \system{} removes nodes to save cost.

Throughout the 32-minute experiment, the latency SLO is satisfied 97\% of the time. 
We first violate the SLO during hot-key replication by 4$\times$ for 15 seconds.
Moreover, the latency spikes to 7$\times$ the SLO during redistribution for about 30 seconds.
Data redistribution causes multicast overhead on the storage servers and address cache invalidation on the clients. 
The latency effects are actually not terrible. 
As a point of comparison, TCP link latencies in data centers are documented tolerating link delays of up to 40$\times$~\cite{Alizadeh:2010:DCT:1851182.1851192}.

From minutes 13 to 18, we meet our SLO of 3.3ms exactly.
With a larger load spike or lower initial resource allocation, \system{} could have easily violated its SLO during that period, putting SLO satisfaction at 83\%, a much less impressive figure.
In any reactive policy, large enough workload variations can cause significant objective violations.
As a result, it is common among cloud providers to develop client-specific service level agreements (SLAs) that reflect access patterns and latency expectations.
In practice, these SLAs allow for significantly more leeway than a service's internal SLO might~\cite{gcpblogpost}.

\subsection{Varying Hotspot}\label{sec:eval-hotspot}

\begin{figure}[t]
  \centering
    \includegraphics[width=0.41\textwidth]{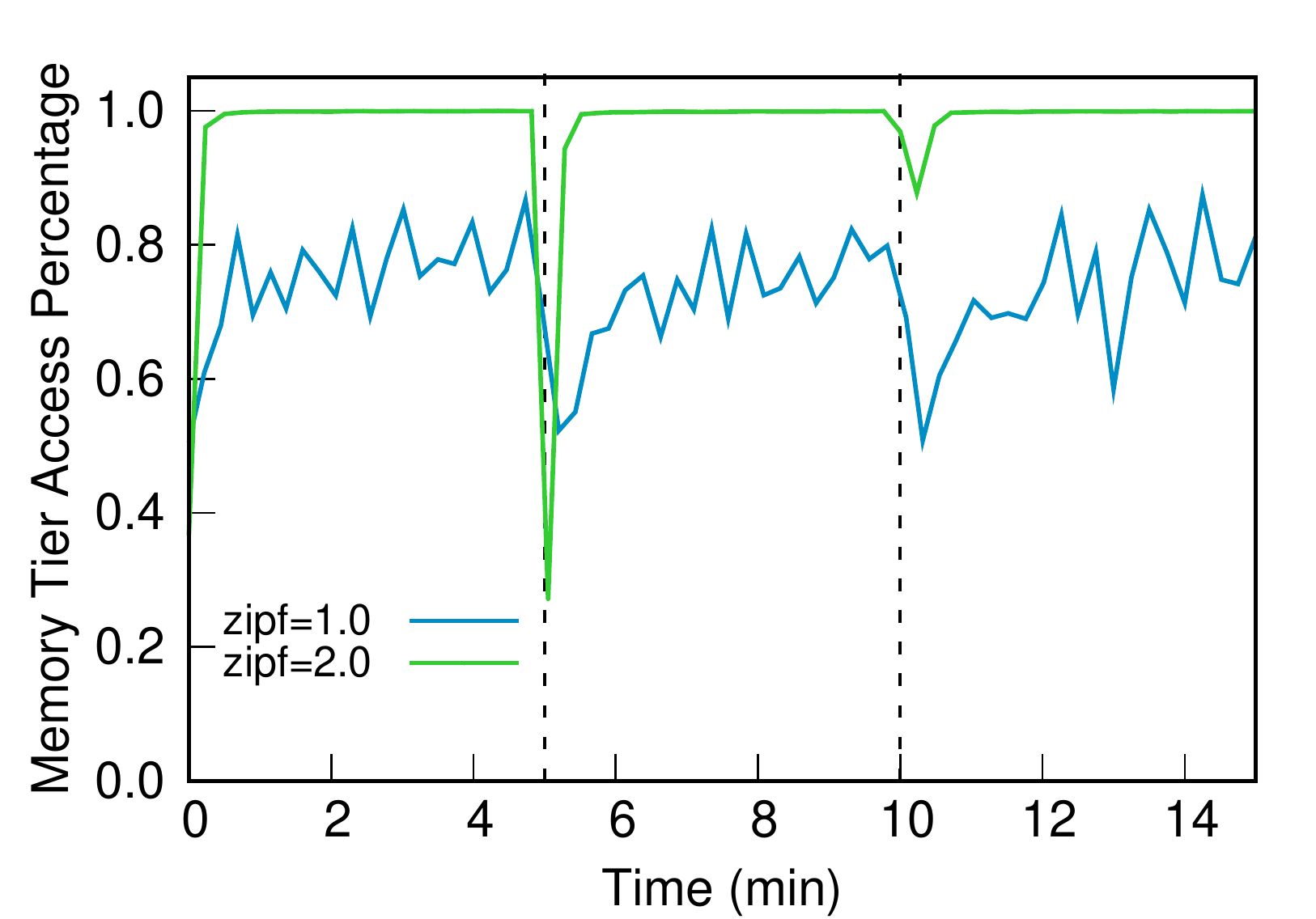}
  \caption{Adapting to changing hotspots in workload.}
  \label{fig:hotspot}
\end{figure}
Next, we introduce multiple tiers and run a controlled experiment to demonstrate the effectiveness of cross-tier promotion \& demotion.
We temporarily disable elasticity.
We evaluate \system{}'s ability to detect and react to changes in workload hotspots.
Here, we do not consider a latency objective, as we are only interested in how quickly \system{} identifies hot data.

We allocate 3 memory nodes (insufficient to store all data) and 15 EBS-tier nodes.
We start with most data residing on the  EBS tier.
The blue curve in Figure~\ref{fig:hotspot} shows a moderately skewed workload, and the green curve shows a highly skewed workload.
At minute 0, we begin a workload centered around one hotspot.
At minute 5, we switch to a different, largely non-overlapping hotspot, and at minute 10, we switch to a third, unique hotspot.
The y-axis measures what percent of queries are served by the memory tier---the ``cache hit'' rate.

We see that for the highly skewed workload, \system{} is able to react to the change almost immediately and achieve a perfect hit rate.
The hot set is very small---on the order of a few thousand keys---and all hot keys are promoted in about ten seconds.
The moderately skewed workload shows more variation.
We see the same dip in performance after the hotspot changes; however, we do not see the same stabilization.
Because the working set is much larger, it takes longer for the hot keys to be promoted, and there is a probabilistic ``fringe'' of keys that are in cold storage at time of access, leading to hit-rate variance.
Nonetheless, we are still able to achieve about an 80\% hit rate less than a minute after the change.

\subsection{Cost-Performance Tradeoffs}\label{sec:eval-trade-off}

\begin{figure}[t]
  \centering
    \includegraphics[width=0.5\textwidth]{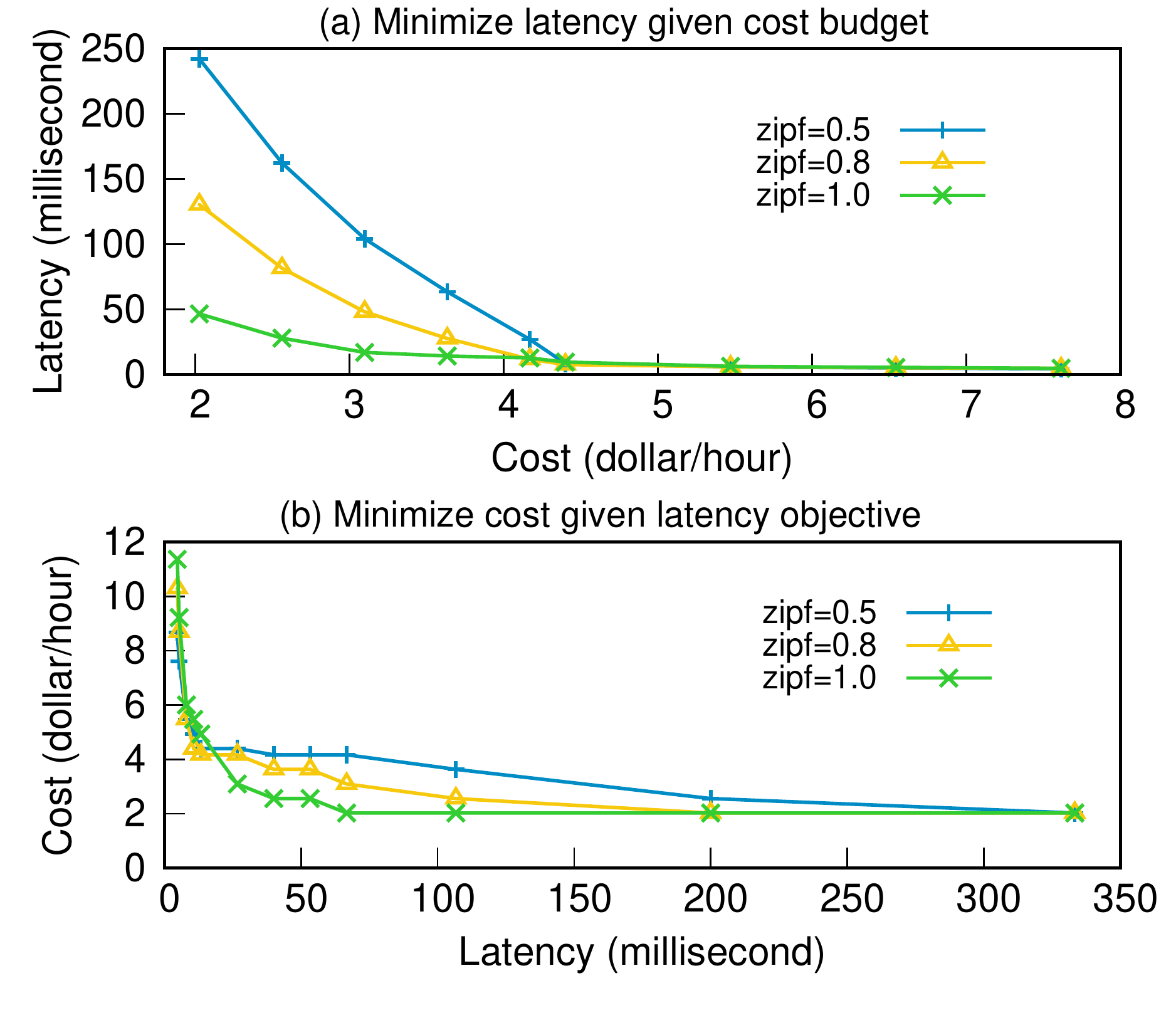}
  \caption{Varying contention, we measure (a) 
  \system{} latency per cost budget; (b) \system{} cost per latency objective.}
  \label{fig:pareto}
\end{figure}

\begin{table}
\centering
{\small
\begin{tabular} { |c|c|c| }
    \hline
    \textbf{Cost} & \textbf{\shortstack{\system{}}} & \textbf{\shortstack{DynamoDB}} \\
    \hline
    \$2.50/hour & 1271 ops/s & 35 ops/s \\
    \hline
    \$3.60/hour & 3352 ops/s & 55 ops/s \\
    \hline
    \$4.40/hour & 23017 ops/s & 71 ops/s \\
    \hline
    \$5.50/hour & 33548 ops/s & 90 ops/s \\
    \hline
    \$6.50/hour & 38790 ops/s & 108 ops/s \\
    \hline
    \$7.60/hour & 43354 ops/s & 122 ops/s \\
    \hline
\end{tabular}
}
\caption{Throughput comparison between \system{} and DynamoDB at different cost budgets.}
\label{table:dynamo}
\end{table}

Finally, we assess how well \system{} is able to meet its SLOs. 
We study the Pareto efficiency of our policy: How well does it find a frontier of cost-performance tradeoffs? 
We sweep the SLO parameter on one of the two axes of cost and latency and observe the outcome on the other.
Here, \system{} uses both storage tiers and enable all policy actions.
We generate workloads with three contention levels---Zipfian coefficients of 0.5 (about uniform), 0.8, and 1.0 (moderately skewed).
For a database of 1M keys with a fault tolerance metric $k=2$, \system{} needs four EBS nodes to store all data and one memory node for metadata; this amounts to minimum deployment cost of \$2.06 per hour.

To ensure that each point was representative and stable, we wait for \system{} to achieve steady state, meaning that nodes are not being added or removed and latency is stable.
In Figure~\ref{fig:pareto}(a), we plot \system{}'s steady state latency for a cost SLO.
We measure average request latency over 30 seconds.
At \$2.10/hour (4 EBS nodes and 1 memory node), only a small fraction of hot data is stored in the memory tier due to limited storage capacity. 
The observed latency ranges from 50ms to 250ms across contention levels.
Requests under the high contention workload are more likely to hit the small set of hot data in the memory tier.
The majority of requests for the low contention workloads hit EBS nodes.
As we increase the budget, latency improves for all contention levels: more memory nodes are added and a larger fraction of the data is memory-resident.
At \$4.40/hour, \system{} can promote at least one replica of all keys to the memory tier.
From here on, latency is under 10ms across all contention levels.
Performance differences between the contention levels are negligible, thanks to hot-key replication.

We also compare the throughput between \system{} and DynamoDB at each cost budget.
Note that in this experiment, DynamoDB is configured to provide the same eventual consistency guarantees and fault tolerance metric ($k=2$) as \system{}.
As shown in Table~\ref{table:dynamo}, \system{} outperforms DynamoDB by 36$\times$ under a low-cost regime and by as much as 355$\times$ at higher costs. 
Our observed DynamoDB performance is actually somewhat better than AWS's advertised performance~\cite{dynamodb-thruput}, which gives us confidence that this result is a reasonable assessment of DynamoDB's efficiency.

Lastly, we set \system{} to minimize cost for a stated latency objective (Figure~\ref{fig:pareto}(b)).
Once more, when the system reaches steady state, we measure its resource cost.
To achieve sub-5ms latency---the left side of Figure~\ref{fig:pareto}(b)---\system{} requires \$9-11 per hour depending on the contention level.
This latency requires at least one replica of all keys to be in the memory tier.
Between 5 and 200ms, higher contention workloads are cheaper, as hot data can be concentrated on a few memory nodes.
For the same latency range, lower contention workloads require more memory and are thus more expensive.
Above 200ms, most data resides on the EBS tier, and \system{} meets the latency objective at about \$2 an hour.

\section{Related Work}\label{sec:related}

As a key-value store, \system{} builds on prior systems, both from the databases and distributed systems literature.
Nonetheless, it is differentiated in the manner in which it leverages and combines these ideas to achieve new levels of efficiency and automation.

\smallitem{Elastic Cloud Storage}.  A small number of cloud-based elastic file systems have considered workload responsiveness and elasticity. 
Sierra~\cite{sierra-practical-power-proportionality-for-data-center-storage} and Rabbit~\cite{Amur:2010:RFP:1807128.1807164} are single-master systems that handle the problem of read and write offloading: when a node is inactive or overloaded, requests to blocks at that node need to be offloaded to alternative nodes. This is particularly important for writes to blocks mastered at the inactive node.
SpringFS~\cite{springfs} optimizes this work by finding a minimum number of machines  needed for offloading. 
By contrast, \system{} supports multi-master updates and selective key replication. When nodes go down or get slow in~\system{}, writes are simply retried at any existing replica, and new replicas are spawned as needed by the policy.

ElastMan~\cite{Al-Shishtawy:2013:EEM:2494621.2494630} is a ``bolt-on'' elasticity manager for cloud KVSes that responds dynamically to changing workload volume. 
\system{}, on the other hand, manages the dynamics of skew and hotspots in addition to volume. 
An interesting aspect of ElastMan is its proactive policy for anticipating workload changes such as diurnal patterns; we return to this point when discussing future work.

\smallitem{Key-Value Stores}. There has been a wide range of work on key-value stores for both multicore and distributed systems---more than we have room to survey here. 
Our earlier work~\cite{anna} offers a recent snapshot overview of that domain. 
In this paper, our focus is not on the KVS kernel, but on the mechanisms to adapt to workload distributions and trade-offs in performance and cost. 

\smallitem{Selective Key Replication}.
The concept of selectively replicating data for performance has a long history, dating back to the Bubba database system~\cite{copeland1988data}. More recently, ecStore~\cite{vo2010towards}, Scarlett~\cite{Scarlett} and E2FS~\cite{Chen2018} perform single-master selective replication, which generate \emph{read-only} replicas of hot data to speed up read performance.
Content delivery network (CDN) providers such as Google Cloud CDN~\cite{gcp}, Swarmify~\cite{swarmify}, and Akamai~\cite{akamai} use similar techniques to replicate content close to the edge to speed up content delivery.
In comparison, \system{}'s multi-master selective replication improves \textit{both} read and write performance, achieving general workload scaling. Conflicts introduced by concurrent writes to different replicas are resolved asynchronously using our lattices' merge logic~\cite{anna}.

Selective replication requires maintaining metadata to track hot keys.  ecStore uses histograms to keep the hot-key metadata compact. \system{} currently maintains access frequencies for the full key set. 
We are exploring two traditional optimizations to reduce overhead in \system{}: heavy hitter sketches rather than histograms~\cite{DBLP:journals/corr/LarsenNNT16}, and the use of distributed aggregation architectures for computing sketches in parallel with minimal bandwidth~\cite{Manjhi:2005:TDE:1066157.1066191}.

Another effort to address workload skew is Blowfish~\cite{Blowfish}, which combines the idea of replication and compression to trade-off storage and performance under time-varying workloads.
Adding compression to \system{} to achieve fine-grained performance cost trade-off is an interesting future direction.

\smallitem{Tiered Storage}.
Beyond textbook caching, there are many interesting multi-tier storage systems in the literature.
A classic example in the file systems domain is the HP AutoRaid system~\cite{tocs96autoraid}. 
Databases also considered tertiary storage  during the era of WORM devices and storage robots~\cite{stonebraker87postgres, sigmod89lomet}. Broadcast Disks envisioned using multiple broadcast frequencies to construct arbitrary hierarchies of virtual storage~\cite{acharya1995broadcast}.
More recently, there has been interest in filesystem caching for analytics workloads. OctopusFS~\cite{Kakoulli:2017:ODF:3035918.3064023} is a tiered file system in this vein.
Tachyon~\cite{Li:2014:TRM:2670979.2670985} is another recent system that serves as a memory cache for analytics working sets, backing a file system interface.
Our considerations are rather different than prior work: The size of each tier in \system{} can change due to elasticity, and the volume of data to be stored overall can change due to dynamic replication. 
\section{Conclusion and Future Work}\label{sec:conclusion}

\system{} provides a simple unified API for efficient key-value storage in the cloud. 
Unlike popular cloud storage systems today, it supports a non-trivial \emph{distribution} of access patterns by eliminating common boundaries in terms of static deployment and cost-performance tradeoffs.
Developers simply declare their desired tradeoffs, without managing a custom mix of heterogenous services.

Behind its simple API, \system{} uses three core mechanisms to meet SLOs efficiently: \emph{horizontal elasticity} to right-size the service by adding and removing nodes dynamically, \emph{vertical data movement across tiers} to reduce cost by demoting cold data, and \emph{multi-master selective replication} to scale request handling at a fine granularity.
The primary contribution of \system{} is its integration of these three features into an efficient, elastic system representing a new design point for cloud storage.
These features are implemented by a policy engine which monitors workloads and responds to them by manipulating metadata stored in \system{}'s high-performance storage engine.

Our evaluation shows that \system{} is extremely efficient. 
In many cases, \system{} is orders of magnitude more cost-effective than popular cloud storage services and prior research systems. 
\system{} is also unique in its ability to automatically adapt to variable workloads. 

\centerline{\rule{1cm}{0.4pt}}\label{sec:future-work}

Although \system{}'s design addresses the main weaknesses of modern cloud storage that we set out to study, it also raises a number of interesting avenues for research.

\smallitem{Proactive Policy Design}. 
Our current policy design is entirely reactive, taking action based on current state. 
To improve this, we are interested in \emph{proactive} policies that anticipate upcoming load spikes and allocate additional resources in advance. 
Using larger workload traces and more advanced predictive techniques, we suspect one could dynamically tune \system{} more intelligently to respond to and anticipate changing workloads.

\smallitem{Defining SLOs \& SLAs}.
Currently, the system administrator defines a single latency objective corresponding to an overall average.
For any system configuration, there are adversarial workloads that can defeat this SLO.
For example, in Section~\ref{sec:eval-dynamic-workload}, a larger load spike could have forced \system{} above its stated SLO for a long period.
SLOs, SLAs and policies can be designed for both expected- and worst-case scenarios, using pricing and incentives.

A fundamental issue is that users with large working sets require more resources at the memory tier to hit a given SLO. 
This is clear in Figure~\ref{fig:pareto}: If each workload corresponds to a user, the user with lower Zipfian parameter costs more to service at a given SLO. 
SLAs should be designed to account for costs varying across users.

\smallitem{Reducing Elasticity Overhead}. The 5-minute delay for node addition noted in Section~\ref{sec:eval-dynamic-workload} is a significant problem. It limits the effectiveness of any elasticity policy, since feedback from allocating a new node is delayed for an eternity in compute time.
Anecdotally, our colleagues building elastic services at major cloud providers tell us they contend with these same issues.
A standard solution today is to maintain a standby pool of ``warm'' nodes that are partially prepared for use. To make this cost-effective, these nodes have to run alternative containers that monetize the idle resources. 
An alternative solution is to make ``cold'' container startup much faster than we experience today. This is a well-studied problem for desktop operating systems~\cite{windowsboot} and VM research~\cite{wood2007black, cully2008remus,lagar2009snowflock}, and we believe it should be more widely available in public cloud settings.

\smallitem{Evaluating Other Tiers}. Currently, \system{} is implemented over only two tiers, but cloud providers like AWS offer a much wider array of price-performance regimes.
There is an opportunity to add services at both ends of the price-performance spectrum that can leverage \system{}'s elastic scaling and coordination-free execution.
As mentioned earlier, our storage kernel requires very little modification to support new storage layers.
Our policy also naturally supports more than two tiers. 
However, our current thresholds in Section~\ref{sec:policy} are the result of significant empirical measurement and tuning.
These parameters will need to be adjusted to the underlying storage hardware.
This manual effort could be replaced by auto-tuning approaches that learn models of configurations, workloads and parameter settings. 
There has been recent work on analogous auto-tuning problems~\cite{van2017automatic,herodotou2011starfish, VanAken:2017:ADM:3035918.3064029}.

%\clearpage
\bibliographystyle{abbrv}
\bibliography{references}
\begin{appendix}
We include pseudocode for the algorithms described in Section~\ref{sec:policy} here.
Note that some algorithms included here rely on a latency objective, which may or may not be specified.
When no latency objective is specified, \system{} aspires to its unsaturated request latency (2.5ms) to provide the best possible performance but caps spending at the specified budget.

\begin{algorithm}[H]
{
    \caption{DataMovement}
    \begin{algorithmic}[1]
    \Require Key, $[<R_M,R_E> <T_M, T_E>]$
        \If {  access(Key, $T$)$>P$ $\&$ $R_M=0$ }
            \State {adjust(Key, $R_M+1$, $R_E-1$, $T_M$, $T_E$)}
        \ElsIf {  access(Key, $T$)$<D$ $\&$ $R_M>0$ }
            \State {adjust(Key, $0$, $k+1$, $1$, $1$)}
        \EndIf
    \end{algorithmic}
    \label{alg:heuristic_movement}
}
\end{algorithm}

\begin{algorithm}[H]
{
    \caption{HotKeyReplication}
    \begin{algorithmic}[1]
    \Require Key, $[<R_M,R_E> <T_M, T_E>]$
        \If {  access(Key, $T$)$>H$ $\&$ $R_M<N_M$ }
            \State {SET $R_{M\_ideal} = R_M*L_{obs}/L_{obj}$}
            \State {SET $R_M' = min(R_{M\_ideal}, N_M)$}
            \State {adjust(Key, $R_M'$, $R_E$, $T_M$, $T_E$)}
        \ElsIf {  access(Key, $T$)$>H$ $\&$ $R_M=N_M$ }
            \State {SET $T_{M\_ideal} = T_M*L_{obs}/L_{obj}$}
            \State {SET $T_M' = min(T_{M\_ideal}, N_{T\_memory})$}
            \State {adjust(Key, $R_M$, $R_E$, $T_M'$, $T_E$)}
        \ElsIf {  access(Key, $T$)$<L$ $\&$ ($R_M>1$ $\|$ $T_M>1$) }
            \State {adjust(Key, $1$, $k$, $1$, $1$)}
        \EndIf
    \end{algorithmic}
    \label{alg:heuristic_replication}
}
\end{algorithm}

\begin{algorithm}[H]
{
    \caption{NodeAddition}
    \begin{algorithmic}[1]
    \Require $tier$, $mode$
        \If {  $mode=storage$ }
            \State {SET $N_{target} =$ required\_storage($tier$)}
                \If { $Cost_{target}>Budget$ }
                    \State {SET $N_{target} =$ adjust()}
                \EndIf
            \State {add\_node(tier, $N_{target}-N_{tier\_current}$)}
        \ElsIf {   $mode=compute$ $\&$ $tier=M$ }
            \State {SET $N_{target} = N_{M\_current}*$min($L_{obs}/L_{obj}$, $c$)}
                \If { $Cost_{target}>Budget$ }
                    \State {SET $N_{target} =$ adjust()}
                \EndIf
            \State {add\_node($M$, $N_{target}-N_{M\_current}$)}
        \EndIf
    \end{algorithmic}
    \label{alg:heuristic_elasticity_add}
}
\end{algorithm}

\begin{algorithm}[H]
{
    \caption{NodeRemoval}
    \begin{algorithmic}[1]
    \Require $tier$, $mode$
        \If {  $mode=storage$ $\&$ $tier=E$ }
            \State {SET $N_{target} =$ max(required\_storage($E$), $k+1$)}
            \State {reduce\_replication()}
            \State {remove\_node($E$, $N_{E\_current}-N_{target}$)}
        \ElsIf {  $mode=compute$ $\&$ $tier=M$ }
            \If {  $N_{M\_current}>1$ }
                \State {reduce\_replication()}
                \State {remove\_node($M$, $1$)}
            \EndIf
        \EndIf
    \end{algorithmic}
    \label{alg:heuristic_elasticity_remove}
}
\end{algorithm}

\begin{algorithm}[H]
{
    \caption{\system{}Policy}
    \begin{algorithmic}[1]
    \Require $tiers = \{M, E\}$, $keys$
        \For {$tier$ in $tiers$}
            \If {  storage($tier$)$>S_{upper}$ }
                \State {NodeAddition($tier$, $storage$)}
            \ElsIf {  storage($tier$)$<S_{lower}$ }
                \State {NodeRemoval($tier$, $storage$)}
            \EndIf
        \EndFor
        \For {$key\in keys$}
            \State {DataMovement($key$)}
        \EndFor
        \If {  $L_{obs}> f_{upper}*L_{obj}$ $\&$ compute($M$)$>C_{upper}$ }
            \State {NodeAddition($M$, $compute$)}
        \ElsIf {  $L_{obs}> f_{upper}* L_{obj}$ $\&$ compute($M$)$<=C_{upper}$ }
            \For {$key\in keys_{memory}$}
                \State {HotKeyReplication($key$)}
            \EndFor
        \ElsIf {  $L_{obs}<f_{lower}*L_{obj}$ $\&$ compute($M$)$<C_{lower}$ }
            \State {NodeRemoval($M$, $compute$)}
        \EndIf
    \end{algorithmic}
    \label{alg:heuristic_policy}
}
\end{algorithm}

\end{appendix}

%\theendnotes

\end{document}